\documentclass{article}
\usepackage{glas}

\usepackage{epsfig}

\newcommand{\qsq}{\ensuremath{Q^2} }

\newcommand{\etbarsq}{\ensuremath{\overline{E_T}^2} }
\newcommand{\xgjets}{\ensuremath{x^{jets}_{\gamma}} }

\newcommand{\xg}{\ensuremath{x_{\gamma}} }

\newcommand{\triple}{\ensuremath{{\rm d}^3\sigma_{ep}/{\rm d}\qsq{\rm
      d}\etbarsq{\rm d}\xgjets} } 
\newcommand{\oos}{$\mathcal{O}(\alpha \alpha_{s}^{2})$}
\def\mb#1{\mbox{\scriptsize {#1}}}
\def\mt#1{\mbox{\tiny {#1}}}

\begin{document}
\begin{titlepage}{GLAS-PPE/1999--03}{March 1999}
\title{Jet Cross Sections at HERA - Current Issues}

\centerline{Steve Maxfield (Dept. of Physics, University of Liverpool),} 
\vspace{0.1cm}
\centerline{Bj\"{o}rn P\"{o}tter (Max-Planck-Institut f\"ur Physik, M\"unchen, Germany)}
\vspace{0.1cm}
\centerline{and Laurel Sinclair (Department of Physics and Astronomy, Glasgow University)}
\vspace{1cm}
\centerline{\it to appear in the proceedings of the 3rd UK Phenomenology Workshop}
\vspace{0.1cm}
\centerline{\it on HERA Physics, September 1998, Durham.}

\vspace{1cm}

\begin{abstract}

Since the start of HERA operation there has been considerable progress
in the understanding of jet production in $e p$ collisions.
QCD calculations are
now able to accommodate the hadronic structure of the virtual photon.
The  luminosities delivered by HERA are now sufficient to allow studies of 
final states in which more than two high transverse energy jets are 
produced.  The transition between jet processes in photoproduction 
and in deep inelastic 
scattering has been studied in some detail.
These advances are highlighted here.

\end{abstract}

\end{titlepage}

\section{Dijet Cross Sections at Low $Q^2$ and Virtual Photon Structure}

\subsection{Comparing Dijet Cross Sections with NLO QCD Calculations}

\begin{figure}[htb]
\begin{center}
\mbox{\epsfig{file=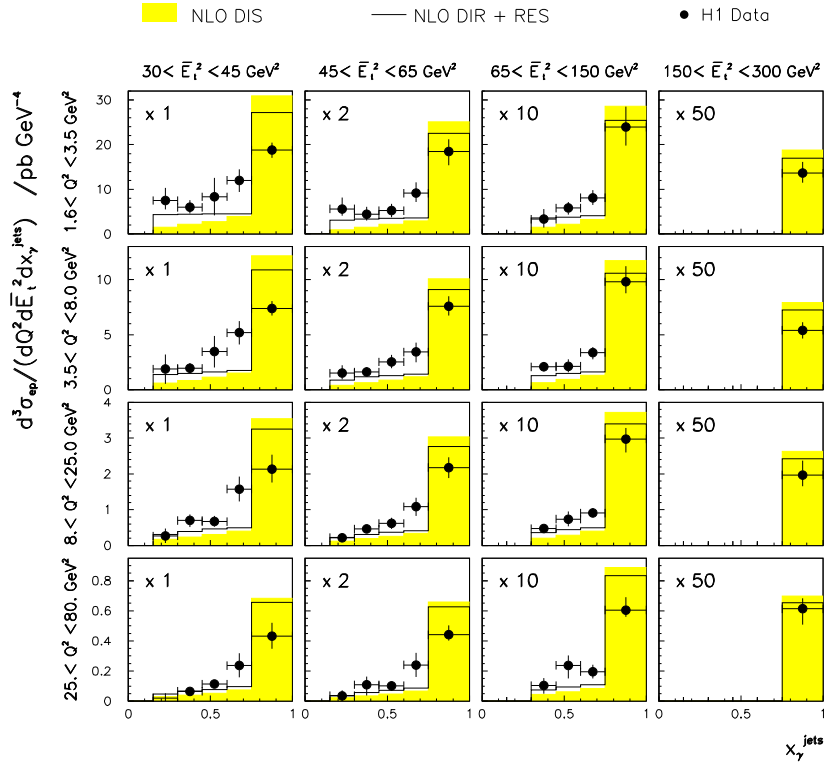,width=11.2cm}}
\end{center}
\vspace{-0.5cm}
\caption{The differential dijet cross-section 
\triple shown as a function of $\xgjets$ for different regions of
$\etbarsq$ and $Q^2$ compared to a NLO QCD calculation which
employed the SaS1M PDF's of the virtual photon. The direct component 
of this model is shown as the shaded histogram. The result including a
resolved component in NLO is shown as the full line.
\label{figure4}}
\end{figure}

H1 have measured the triple-differential cross-section, $\triple$
($\etbarsq$ is the mean $E_T$ of the two highest $E_T$ jets) in the
$\gamma^*p$ center-of-mass system (hadronic cms)~\cite{vanc2}. 
The photon virtuality spans the range $1.6 < Q^2 < 80 {\rm GeV}^2$ 
and $y=Pq/Pk$ is constrained to $0.1<y<0.7$,
where $P$, $q$ and $k$ are the four-vectors of the proton, virtual
photon, and electron. The momentum fraction of the parton from the photon 
entering the hard scattering, $\xgjets$, defined as 
\begin{equation} \label{xgam}
 \xgjets = \frac{\sum_{i=1,2} E^{jets}_{T_i}\exp
 (-\eta_i^{jets})}{W} 
\end{equation}
is estimated from the two highest $E_T$ jets. The variable
$W^2=(P+q)^2=2Pq -Q^2$ defines the hadronic cms energy.\footnote{One 
can theoretically define the variable
$x_\gamma=p_0P/qP$, where $p_0=x_\gamma q$ is 
the four-vector of the incoming parton from the photon. This
definition assumes a collinear emission of the parton from the photon,
which is an approximation neglecting some $k_\perp$ contribution
due to the finite $Q^2$. These contributions are, however, small for
moderate $Q^2$. This variable differs from that defined in
(\ref{xgam}). When comparing with QCD calculations at LO level,
partons directly give jets and there are exactly two partons in the
final state. Thus, $x_\gamma^{jets} = x_\gamma$. However, in NLO,
$x_\gamma$ still gives the momentum fraction of the parton in the
photon, but $x_\gamma^{jets} \ne x_\gamma$. At NLO, one also has
contributions to $x_\gamma^{jets} <1$ from the direct contribution,
whereas at LO $x_\gamma^{jets}=1$ for the direct process.}
The triple differential cross-section is shown in figure~\ref{figure4} 
as a function of $\xgjets$ in ranges of
$Q^2$ and $\etbarsq$. The data (points) are corrected for detector
effects and the error bar shows the quadratic sum of systematic and
statistical errors. The cross section decreases rapidly with
increasing $\etbarsq$ and with increasing $Q^2$. Hadronization
corrections are not included in the data. These are believed to be
small on average but will presumably change the shape of the $\xgjets$
distribution. 

QCD calculations of dijet cross-sections with virtual
photons have recently been performed at next-to-leading order (NLO)
\cite{1,2}. The calculations are implemented in the fixed order
program {\tt JetViP} \cite{jv}. In NLO, in the direct component a
logarithm $\ln E_T^2/Q^2$ occurs, which is proportional to the photon
splitting function. This term is large for $E_T^2\gg Q^2$ and
therefore subtracted and resummed in the virtual photon structure
function. The condition $E_T^2\gg Q^2$ ensures that it is
possible to resolve an internal structure of the virtual photon. The
virtual photon PDF's are suppressed as $Q^2 \rightarrow E_T^2$ and
various anz{\"a}tze have been used to interpolate between the regions
of known leading-log behaviour~\cite{SAS,vpth2,vpth5,vpth6}. It should
be noted that the logarithmic term is only subtracted for the 
transversely
polarized photons, since it vanishes in the case of longitudinally
polarized photons for $Q^2\to 0$. The hadronic content of longitudinal
virtual photons should be very small and therefore negligible.
The results of the NLO calculations are also shown in
figure~\ref{figure4}. The direct component of this model is shown as 
the shaded histogram. The result including a resolved component in NLO
is shown as the full line. The calculation including a resolved photon
component compares better to the data and indicates a need for 
a resolved virtual photon component below $10$~GeV$^2$, especially in
the forward rapidity region, corresponding to low $\xgjets$. The
discrepancy of data and NLO calculation for $\xgjets >0.75$ will
presumably be cured when hadronization effects are considered, which
lower the NLO results at large $\xgjets$.

\subsection{Effective Virtual Photon Parton Densities}

\begin{figure}[htb]
\begin{center}
\mbox{\epsfig{file=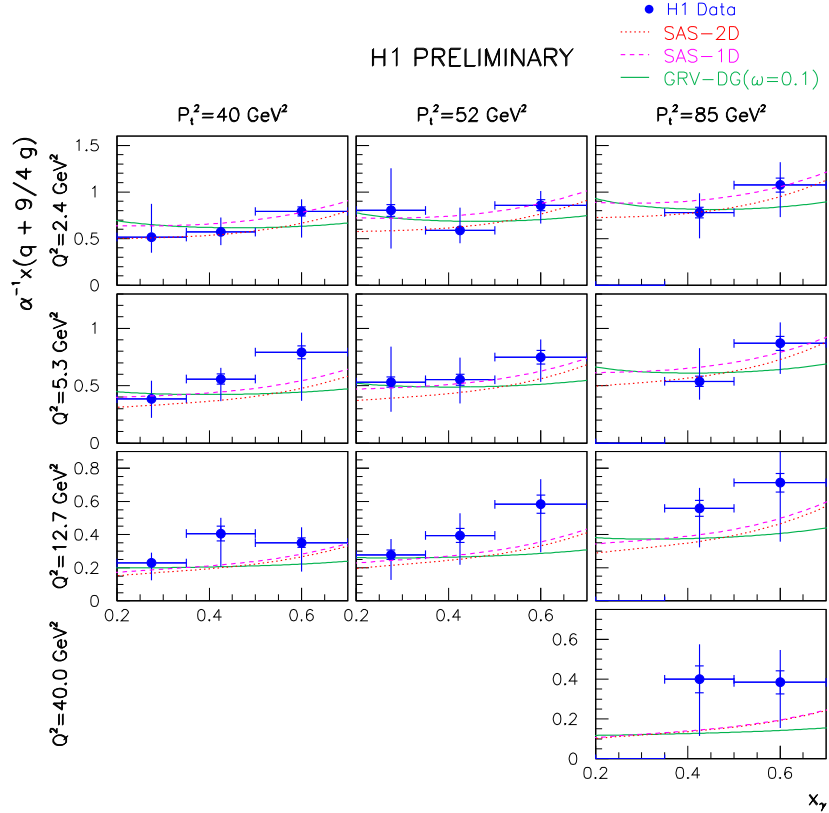,width=10.4cm}}
\end{center}
\caption{The leading order effective parton density of the photon 
$x(q + 9/4 g)$, divided by the fine structure constant $\alpha$,
as a function of $x_\gamma$ 
for different values of $Q^2$ and the transverse momentum $P_t^2$. 
The data are displayed as points, with the inner error bar depicting
the statistical error, and the total error bar the quadratic
sum of statistical and systematic errors. Also shown are
the prediction from the DG model using GRV-LO real photon parton 
densities and $\omega=0.1$ (solid line) and the SAS-1D (dashed line) 
and SAS-2D(dot-dashed line) parameterisations. 
\label{figure6}}
\end{figure}

H1 have used their studies to extract an effective parton denstity
(EPDF) of the virtual photon. By using the Single Effective
Subprocess Approximation~\cite{SES}, the cross-section for dijet
production in LO can be written
\[ \!\!\!\!\!\!\!\!\!\!\!\!\!\!\!\! \frac{{\rm d}^5\sigma}{{\rm d}y{\rm d}x_{\gamma}
   {\rm d}x_{\rm p}{\rm d}\cos\theta^*{\rm d}Q^2} \sim 
   \frac{f_{eff/\gamma}^k(x_\gamma,P_t^2,Q^2)}{x_\gamma}
   \frac{f_{eff/{\rm p}}(x_{\rm p},P_t^2)}{x_{\rm p}}
   |M_{SES}(\cos\theta^*)|^2 . \]
Here the Effective Parton Densities (EPDF), $f_{eff/\gamma}^k$ and
$f_{eff/{\rm p}}$, are defined as 
\begin{math}
f_{eff/{\rm A}} \equiv (f_{q/A}+f_{\overline{q}/A}) + \frac{9}{4} f_{g/A}.
\end{math}
The EPDF is shown in figure~\ref{figure6}. 
We see that, independently
of $Q^2$, the parton density tends to be flat or  
rising with $\xg$ (not to be confused with $\xgjets$). This behaviour
is maintained as the probing scale increases. These are features
characteristic of photon structure. The data are compared with
predictions from the SaS~\cite{SAS} and DG~\cite{vpth5,vpth6} models which
are able to describe the data quite well except where $Q^2 \to P_t^2$
and various aspects of the model start to break down. The 
models tend to underestimate the data in these regions.  The three
parameterisations for the parton density all give a good description
of the data both in the lowest $\xg$ range and in the lowest
two $Q^2$ bins but predict a more rapid suppression as $Q^2\to P_t^2$ 
than is seen in the data.

\section{Three Jet Photoproduction}

\begin{figure}[htb]
\begin{center}
\mbox{\epsfig{file=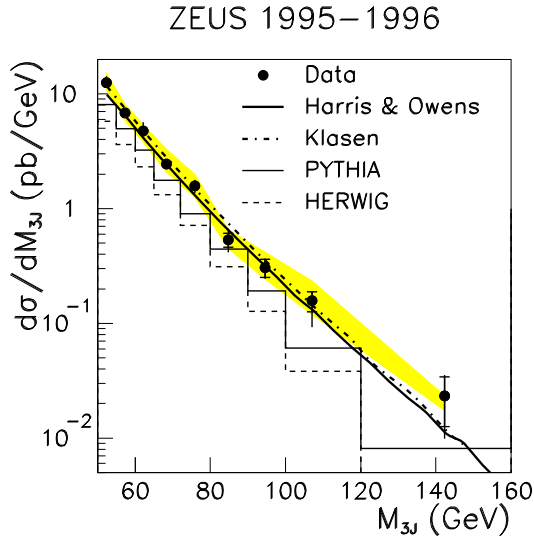,width=7.cm}}
\end{center}
\caption{The three-jet cross section $d\sigma/dM_{\mt{3J}}$.
The dots show the data.
The inner error bars show the statistical error.  The outer error bars
show the quadratic sum of the statistical and systematic uncertainties
with the exception of the absolute energy scale uncertainty which
is shown separately as a shaded band.
\oos\ pQCD calculations
by two groups of authors are shown by thick solid and dot-dashed lines.
The thin solid and dashed histograms show the predictions from PYTHIA
and HERWIG.
\label{M3J}}
\end{figure}

ZEUS has measured the high-mass three-jet cross section in 
photoproduction, $d\sigma/dM_{\mb{3J}}$, as shown in
Figure~\ref{M3J}~\cite{ZEUS_3J}.
The \oos\ pQCD calculations from two groups of 
authors~\cite{harris,klasen} provide a good description of the data, even 
though they
are leading order for this process.  Monte Carlo models also generate
three-jet events through the parton shower mechanism and both
PYTHIA~\cite{PYTHIA} and HERWIG~\cite{HERWIG} reproduce the shape of 
the $M_{\mb{3J}}$ distribution.

\begin{figure}[htb]
\begin{center}
\mbox{\epsfig{file=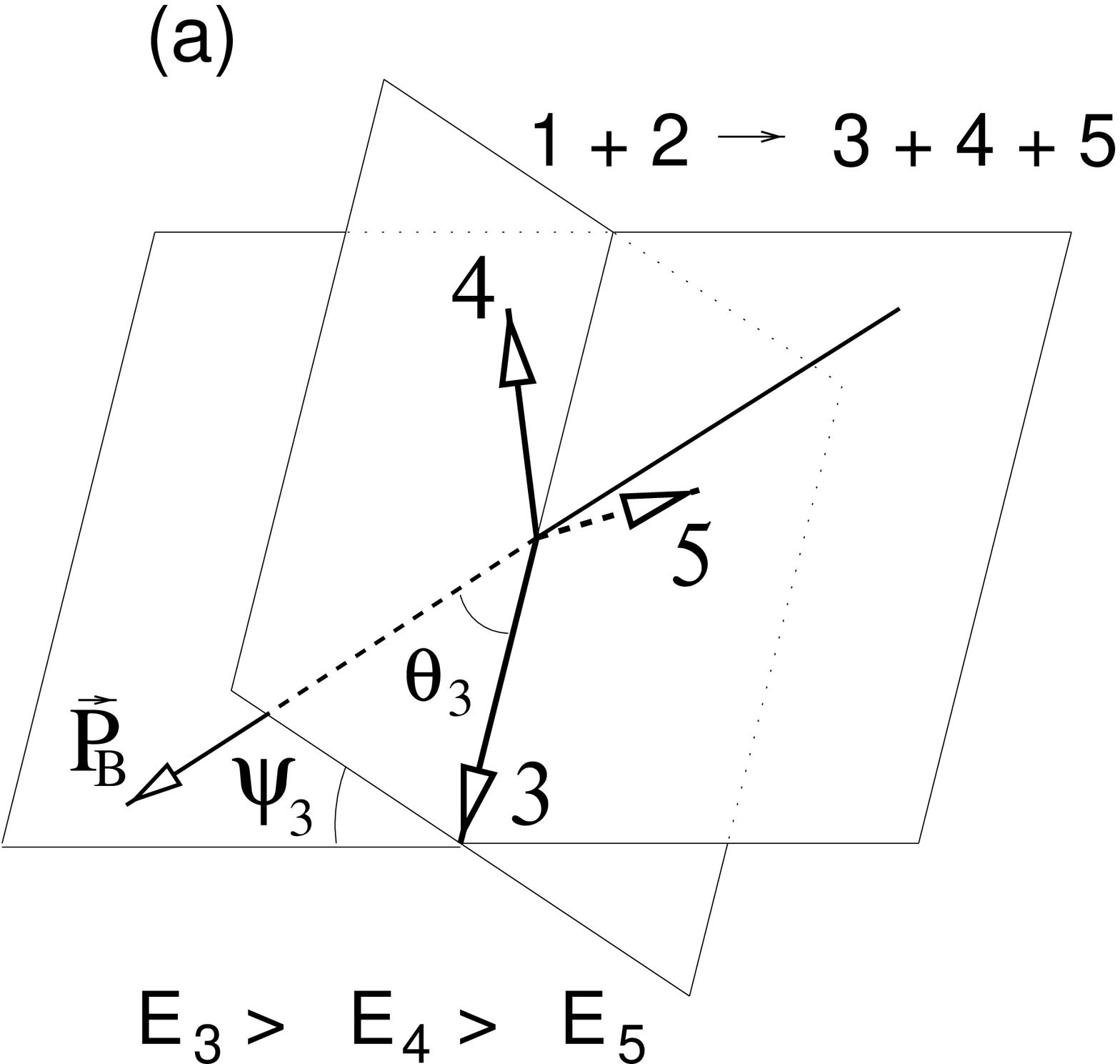,width=4.6cm}}\mbox{\epsfig{file=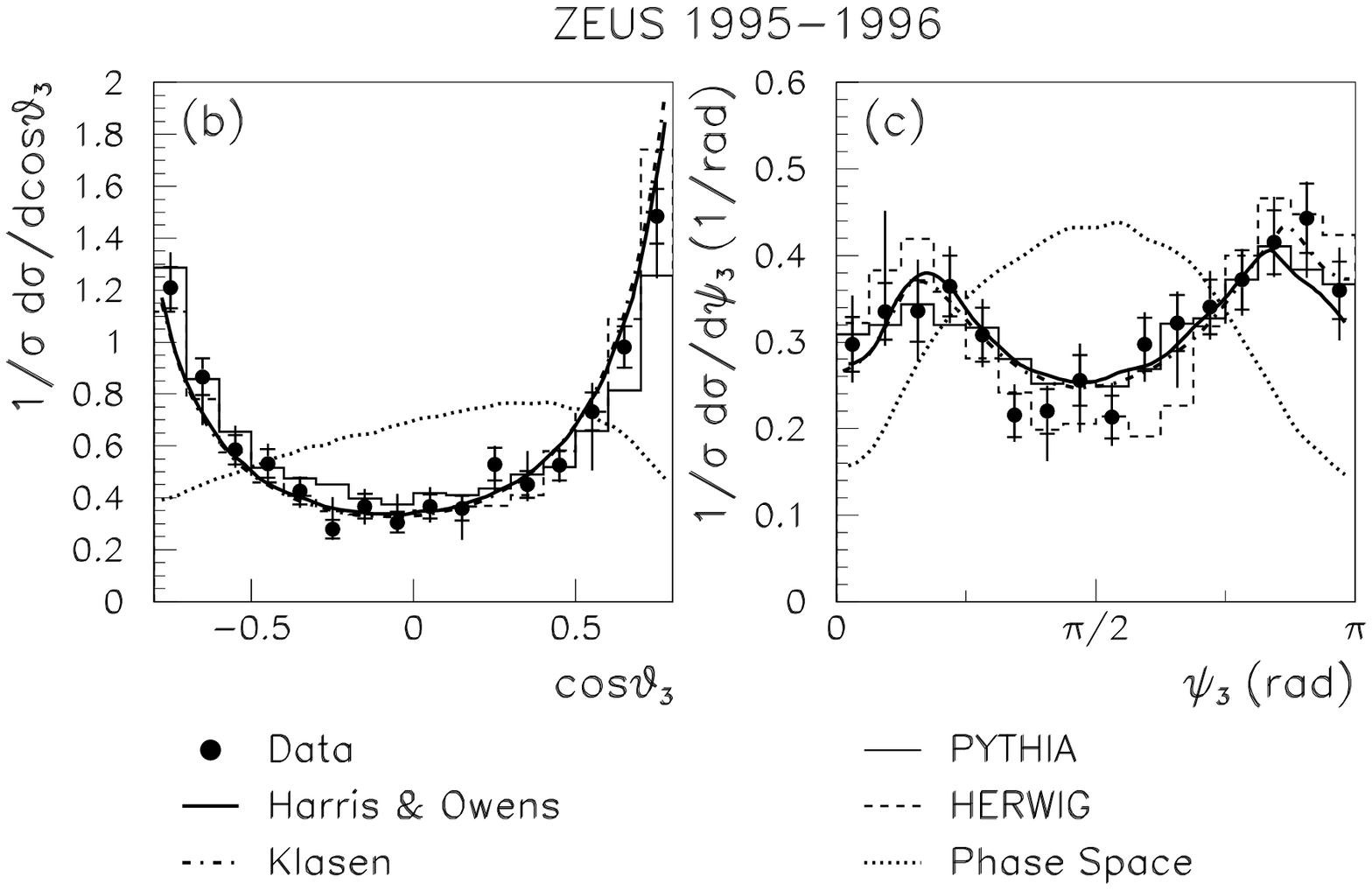,width=9.cm}}
\end{center}
\caption{(a) Centre-of-mass frame diagram of the three-body scattering angles.
The beam direction is indicated by $\vec{P}_{B}$.  Distributions  
of $\cos \theta_3$ and $\psi_3$ are shown in (b) and (c).  The dotted curve 
shows the 
distribution for a constant matrix element.  Other details are as described 
for Figure~\ref{M3J}.
\label{angles}}
\end{figure}

For three-jet events there are two relevant scattering angles as
illustrated in Figure~\ref{angles}(a).  
The distributions of
$\cos \theta_3$ and $\psi_3$ are shown in
Figure~\ref{angles}(b) and (c).
The $\cos \theta_3$ distribution resembles that of $\cos \theta^*$ in dijet
production~\cite{ZEUS_2J} and exhibits forward and backward peaks. 
It is well described in both
\oos\ pQCD calculations and parton shower models.  The $\psi_3$ 
distribution is peaked near 0 and $\pi$ indicating that the three-jet
plane tends to lie near the plane containing the highest energy jet
and the beam.  This is particularly evident if one considers the
$\psi_3$ distribution for three partons uniformly distributed in the
available phase space.  The phase space near $\psi_3 = 0$ and $\pi$
has been depleted by the $E_T^{\mb jet}$ cuts and by the jet-finding
algorithm.  The pQCD calculations describe perfectly the $\psi_3$
distribution.  It is remarkable that the parton shower models PYTHIA
and HERWIG are also able to reproduce the $\psi_3$ distribution.

Within the parton-shower model it is possible to determine the
contribution to three-jet production from initial-state radiation (ISR)
and final-state radiation.  It is also possible to switch the QCD phenomenon
of colour coherence on and off.  From a Monte Carlo study it has been 
determined that ISR is predominantly responsible for three-jet production.
Also, it has been found that colour coherence can account for the 
suppression of large angle emissions which leads to the depletion of the 
$\psi_3$ distribution near $\psi_3 = \pi / 2$~\cite{ZEUS_3J}.

\section{Summary}

The dijet cross section at low $Q^2$ has been measured by the H1
collaboration and compared to NLO QCD calculations. The comparison
shows a clear need for resolved virtual photon component. First
measurements of a leading order effective virtual photon PDF have been
made. Existing models for virtual photon PDF's are consistent
with the measurements but systematic errors are still large.

A first measurement of high transverse energy three-jet 
photoproduction has been performed.  The distribution of the three 
jets is sensitive to colour coherence and is correctly predicted in 
\oos\ pQCD.


\end{document}